\documentclass{article}
\usepackage{mytex}
\usepackage{eepic}

\begin{document}
 
\centerline{\Large\bf Sensitivity of a Shallow-Water Model to Parameters.}
\vspace{5mm}

\centerline{\Large \bf Eugene Kazantsev}

\begin{center}
 INRIA, projet MOISE, 
 Laboratoire Jean Kuntzmann,\\
BP 53,
38041 Grenoble Cedex 9, 
France \\\begin{tabular}{ll}
Telephone:&+33 4 76 51 42 65 \\
Fax:& +33 4 76 63 12 63\\
E-Mail:&kazan@imag.fr
\end{tabular} 
\end{center}
\vspace{5mm}
{\bf Abstract:}
{\small
An adjoint based technique is applied to a shallow water model in order to estimate the  influence of the model's parameters on  the solution. Among parameters  the bottom topography, initial conditions,   boundary conditions on rigid boundaries, viscosity coefficients Coriolis parameter and the amplitude of the wind stress tension are considered.  Their influence is analyzed from three points of view:
 \begin{itemize}
\item flexibility of the model with respect to a parameter that is related to the lowest value of the cost function that can be obtained in the data assimilation experiment that controls this parameter;
\item possibility to improve the model by the  parameter's control, i.e. whether the solution with the optimal parameter remains close to observations  after the end of  control;
\item sensitivity of the model solution to the parameter in a classical sense.  That implies the analysis of the sensitivity
estimates and their comparison with each other and with the local Lyapunov exponents that characterize the sensitivity of the model to initial conditions. 
\end{itemize}
Two configurations have been analyzed: an academic case of the model in a square box and a more realistic case simulating   Black Sea currents. It is shown in both experiments that  the boundary conditions near a rigid  boundary influence the most the solution. This fact points out the  necessity to identify optimal boundary approximation during a model development.

}

{\bf Keywords:}
{ Variational Data Assimilation; Sensitivity to parameters; Boundary conditions; Shallow water model.}



\section{Introduction}

 Lorenz, in his  pioneering work  \cite{Lor63} has shown that a geophysical fluid is extremely sensitive to initial conditions and  perturbations of the initial state grow exponentially in time. This discovery led to the development of the sensitivity studies   intended to  describe the evolution of a small unknown error in initial data and its influence on the forecast. 
 
 Together with the sensitivity studies,  data assimilation methods have also  been under rapid development. These methods are intended to bring  the model and various observational information together in order to better identify the initial state of the model.

The variational data assimilation technique,  first proposed in \cite{Ledimet82}, \cite{ldt86}, is based on the optimal control methods \cite{Lions68} and perturbations theory   \cite{Marchuk75}. This technique allows us to retrieve an optimal data for a given model from heterogeneous observation fields. Since the early 1990s, many mathematical and geophysical teams are involved in the development of the data assimilation strategy. One can cite many papers devoted to this problem, as  in the domain of development of different methods for the  data assimilation  and also in the domain of its applications to the atmosphere and oceans.  

In the beginning, data assimilation methods were intended to identify and reconstruct an optimal initial state for the model. However, the idea that other model's parameters should  also be identified by data assimilation has also been studied and discussed in numerous papers.  
One can cite several examples of using  data assimilation  to identify the bottom topography of simple models (\cite{LoschWunsch}, \cite{Heemink}, \cite{assimtopo}), to control open boundary conditions in coastal and regional models (\cite{shulman97}, \cite{shulman98}, \cite{Taillandier}, \cite{Brummelhuis}), boundary conditions on rigid boundaries (\cite{fxld-mo}, \cite{Leredde}  \cite{Lellouche}, \cite{assimbc1}, \cite{sw-lin}, \cite{sw-nl})  and to determine  other parameters of a model (\cite{zou}, \cite{panchang}, \cite{chertok}). 
 
This paper is devoted to the comparison  and  the analysis of the dependence of the solution of a shallow-water model on different parameters. All available model's parameters have been included in the test set. First of all, among these parameters we consider the initial state of the model in order to compare the influence of all other model's parameters with the influence of the initial state. Second, we study the influence of the boundary conditions on rigid boundaries. But, instead of considering the boundary conditions themselves, we  remind that particular attention must be paid to the discretization process (see \cite{Leredde}) that introduces the boundary conditions into the model. So,  we  use the representation  proposed in \cite{assimbc1} that allows us to  control  the discretization of the model's operators near the boundary and to obtain  immediately the numerical scheme that deal with the boundary conditions.  Third, we also include  the bottom topography in the set of  control parameters because the importance of optimal representation of the  topography on the model grid has been discussed in numerous  papers, as  cited above and others.  In addition, the control set includes also several scalar parameters like dissipation coefficients and forcing amplitude.  All these parameters have  empirical origins, their values can also be optimized. And finally, we control the Coriolis parameter. 

The influence of the parameters on the model solution is analyzed from three points of view. First, we compare the flexibility of the model with respect to a particular parameter. The flexibility in this paper is understood as the capability of a parameter to bring the model's solution closer to observations reducing the cost function value in the assimilation procedure. Of course, observational data contain measurement errors and result from the different physical processes in the nature. Consequently, there is no hope to get vanishing cost function varying any model's parameter. However, if variation of some parameter allows to bring the model's solution closer to observations, we can consider the model is more flexible with respect to this parameter.

However, flexibility of the model cannot be considered as a real improvement of the model's solution. Low cost function's value is obtained in frames of  the minimization procedure, under strong external control that force the model solution toward observational data. The distance between the solution and observations can rapidly increase after the end of the assimilation when the model is no longer forced to remain close to observations.  That is why we analyze  also the behavior of the solution beyond the assimilation window and examine  the distance on longer time scales  considering that data assimilation improves the model if the solution remains close to observations after the end of control. 

And the third sensitivity estimate considered in this paper is similar to the classical sensitivity characteristic that relates  the  norm of the perturbation of the solution and the norm of  the perturbation of the parameter. Local (or finite time)  Lyapunov exponents, for example, are evaluated using this ratio with  initial conditions used as a parameter.  This ratio shows how much a small normalized error in a parameter perturbs the solution at a given time.

  Two examples are considered in this paper: an academic case of the model in a  square box with artificially generated observations and a more realistic case of  assimilation of   real observational data in the Black sea model. We should note that the same model is used in both configurations, but  the  difference  consists in chaotic behavior of the model in the square box and regular flow in the Black sea. Temporal variability of the solution in the last case is only due to variations of the wind stress on the sea surface.

\section{Shallow Water Model}

\subsection{Model's equations and discretization}
\label{sec1}

In this paper we consider   a  shallow-water model   written in a conservative form:
\beqr
\der{hu}{t}&=&  f hv - \der{}{x}\biggl( hu^2+gh(h-H)-\mu h\der{u}{x}\biggr)- 
 \nonumber \\&-&
  \der{}{y}\biggl(h u v-\mu h\der{u}{y}\biggr) 
   -\sigma hu+\tau_0 \tau_x, 
\nonumber \\
\der{hv}{t}&=&-f hu - \der{}{x}\biggl( h u v-\mu h\der{v}{x}\biggr)+
 \nonumber \\&-&
  \der{}{y}\biggl(\ hv^2+gh(h-H)-\mu h\der{v}{y}\biggr) 
    -\sigma hv  +\tau_0 \tau_y, 
\label{sw}  \\
\der{h}{t} &=& -\der{h u}{x}-\der{h v}{y}. \nonumber
\eeqr
where $hu(x,y,t)$ and $hv(x,y,t)$ are  two flux components that represent the product of the velocity by the ocean depth, $h(x,y,t)$, that corresponds to the distance from the sea surface to the bottom of the ocean. The sea  surface elevation is represented by the difference $h(x,y,t)-H(x,y)$, where $H(x,y)$ is the bottom topography.  The model is driven by the surface wind stress with components $\tau_x(x,y,t)$ and $\tau_y(x,y,t)$ normalized by $\tau_0$ and subjected to the  bottom drag that is parametrized by  linear terms $\sigma hu$ and $\sigma hv$. Horizontal eddy diffusion is represented by   harmonic operators  $div( \mu h \nabla u)$ and $div( \mu h \nabla v)$. Coriolis parameter is represented by the variable $f(y)$ that is equal to $f_0+\beta y$ assuming $\beta$-plane approximation. Parameter   $g$ is the reduced gravity.  The system is defined in some domain $\Omega$ with characteristic size $L$  requiring that   both  $hu$ and $hv$ vanish on the whole boundary of $\Omega$. No boundary conditions are prescribed for $h$.  Initial conditions are defined for all variables: $hu,\; hv$ and $h$. 

As usual, initial conditions are considered as the control parameter of the model in this paper. We study the sensitivity of the model to its initial point and assimilate data to find its optimal value. However, in addition to initial conditions, 
all other parameters of the model, and namely its bottom topography $H(x,y)$, Coriolis parameter $f=f(y)$, scalar coefficients $\mu, \sigma, g$ and $\tau_0$,  are also considered as control variables. All of them are allowed to vary in the data assimilation procedure in order to bring them to their optimal values. The sensitivity of the model with respect to small variations of these parameters will also be studied. 

As it has been shown in  \cite{sw-nl}, \cite{sw-lin},  the influence of the boundary conditions on rigid boundaries on the model's solution is also strong. On the one hand, assimilating data allows us to  better represent the model's boundary  on the model's grid, and, on the other hand, it allows to correct several types of errors committed by numerical schemes and finite dimensional approximations of the model's operators.  
 
However, following   \cite{sw-lin}, instead of controlling boundary conditions, we choose to control the way they are introduced into the model. And namely, we consider the discretization of the model's operators  near the boundary as the control parameter, that means the numerical scheme that takes into account the set of boundary conditions. A detailed description of controlling the discretization as well as the analysis of the tangent and adjoint models can be found in   \cite{sw-nl}. Here, we just remind the principal points.

We discretize all variables of this equation on the regular  Arakawa's C-grid \cite{AL77} with constant grid step $\delta x=\fr{L}{N}$ in both $x$ and $y$ directions:
\beqr
hu_{i,j-1/2}(t)&=&hu(i h,j h-h/2,t) \mbox{ for } i=0,\ldots N, j=0,\ldots,N+1 \nonumber \\
hv_{i-1/2,j}(t)&=&hv(ih-h/2,jh,t) \mbox{ for } i=0,\ldots N+1, j=0,\ldots N \nonumber \\
h_{i-1/2,j-1/2}(t)&=&h(ih-h/2,jh-h/2,t) \mbox{ for } i=0,\ldots N+1, j=0,\ldots N+1 \nonumber 
\eeqr

Discretizing  system \rf{sw}, we replace the derivatives by their finite difference representations  $D_x$ and $D_y$ and   introduce  two    interpolations in $x$ and $y$ coordinates $S_x$ and $S_y$. Interpolations are necessary on the staggered grid  to calculate the variable's values in nodes where other variables are defined.  The discretized system \rf{sw} writes

\beqr
\der{hu}{t}&-& f S_xS_yhv+
D_x\biggl(\fr{(S_x hu)^2}{h}+gh(h-H(x,y))-\mu h D_x\fr{hu}{S_xh}\biggr)+ 
 \nonumber \\&+&
  D_y\biggl(\fr{(S_y hu\; S_x hv)}{S_x S_y h}-\mu (S_x S_y h)D_y\fr{hu}{S_x h}\biggr) 
  =  -\sigma hu+\tau_0\tau_x, 
\nonumber \\
\der{hv}{t}&+& f S_yS_x hu +D_x\biggl(\fr{(S_y hu\; S_x hv)}{S_x S_y h}-\mu (S_xS_y h) D_x\fr{hv}{S_y h}\biggr)+ 
 \nonumber \\&+&
  D_y\biggl(\fr{(S_y hv)^2}{h}+gh(h-H(x,y))-\mu h D_y\fr{hv}{S_y h}\biggr) 
  =  -\sigma hv  +\tau_0\tau_y, 
\label{sw-grid}  \\
\der{h}{t} &=&- D_x hu - D_y hv. \nonumber
\eeqr
Discretized operators $D_x, D_y$ and $S_x, S_y $ are defined in a classical way at all internal points of the domain. For example,  the second order derivative and the interpolation operator of the variable $hu$ defined at corresponding  points  write
\beqr
(D_x hu)_{i-1/2,j-1/2}&=&\fr{hu_{i,j-1/2} -hu_{i-1,j-1/2} }{\delta x}  \mbox{ for } i=2,\ldots , N-1,
 \nonumber \\
(S_x hu)_{i-1/2,j-1/2}&=&\fr{hu_{i,j-1/2} +hu_{i-1,j-1/2} }{2}\mbox{ for } i=2,\ldots , N-1.  \label{intrnlsch} 
\eeqr
To calculate fourth order approximations of derivatives and interpolations we use the following formulas 
\beqr
(D_x hu)_{i-1/2,j-1/2}&=&\fr{hu_{i-2,j-1/2} -27hu_{i-1,j-1/2}+27 hu_{i,j-1/2} -hu_{i+1,j-1/2} }{24\delta x},
 \nonumber \\
(S_x hu)_{i-1/2,j-1/2}&=&\fr{-hu_{i-2,j-1/2} +9hu_{i-1,j-1/2}+9hu_{i,j-1/2} -hu_{i+1,j-1/2} }{16}.  \label{intrnlsch4} 
\eeqr

   Discretization of operators in the  directly adjacent to the boundary nodes are  different from \rf{intrnlsch} and represent the control variables in this study. In order to obtain their  optimal values  assimilating external data, we suppose nothing about derivatives and interpolations near the boundary  and  write them in a general form
\beqr
(D_x hu)_{1/2,j-1/2}&=&\alpha_{0}^{D_x^{hu}}+\fr{\alpha^{D_x^{hu}}_{1} hu_{0,j-1/2} +\alpha^{D_x^{hu}}_{2} hu_{1,j-1/2}}{\delta x} \nonumber\\
(S_x hu)_{1/2,j-1/2}&=&\alpha_{0}^{S_x^{hu}}+\fr{\alpha^{S_x^{hu}}_{1} hu_{0,j-1/2} +\alpha^{S_x^{hu}}_{2} hu_{1,j-1/2}}{2}
 \label{bndsch}
\eeqr

This formula represents a linear combination of values of $hu$ at two  points adjacent to the boundary with  coefficients $\alpha$. The constant $\alpha_0$ may be added in some cases to simulate non-uniform boundary conditions like $hu(0,y)=\alpha_0\neq 0$.

We distinguish $\alpha$ for different variables and different operators allowing different controls of  derivatives   because of the different nature of these variables and different boundary conditions prescribed for them. It is obvious, for example, that the approximation of the derivative $D_x$ in the first equation may differ from the approximation of  $D_x$ in the third one. Although both operators represent a derivative,  boundary conditions for $hu$ and $h$ are different; these derivatives are defined at different points, at different distance from the boundary. Consequently, it is reasonable to let them be controlled separately and  to assume that their optimal approximation may be different with distinct coefficients $\alpha^{D_x^{hu}}$ and $\alpha^{D_x^{h}}$.

Time stepping of this model is performed by the leap-frog scheme. 
The first time step is split into two Runge-Kutta stages in order to ensure the second order approximation.

The approximation of the derivative introduced by \rf{intrnlsch} and \rf{bndsch} depends on variables $\alpha$. They are added to the set of control variables enumerated above. Operators  are allowed to change their properties near boundaries in order to find the best fit with requirements of the model and data.   To assign all control variables   we shall perform the data assimilation procedure and find their optimal values.  Variational data assimilation is usually performed by minimization of the specially introduced cost function. The minimization is achieved using the gradient of the cost function that is usually determined by the run of the adjoint to the tangent linear model.

\subsection{Cost function}

One of the principal  purposes of variational data assimilation consists in the variation of  control parameters in order to bring the model's solution closer to the observational data. This implies the necessity to measure the distance between the trajectory of the model and the data. Introducing the cost function, we define this measure. Generally speaking, the cost function is represented by some norm of the difference between the model's solution and observations, eventually accompanied by some regularization term.

 To define the cost function we introduce dimensionless  state vector $\phi$ that is composed of three variables of the model $\phi=\{w_{hu}hu,w_{hv}hv,w_h h\}^t$ weighted by coefficients $w$. These weights  are used to normalize values of the flux components by  $w_{hu}=w_{hv}=\fr{1}{H_0\sqrt{gH_0}}$ and the Sea surface elevation by $w_h=\fr{1}{H_0} $. The distance between the model solution and observations is defined as the Euclidean norm of the difference

\beqr
\xi^2&=&\xi^2(\phi(p,t))=\sum_{k} (\phi_{k}- \phi^{obs}_{k})^2 =
\label{xiphi}\\
&=&w_{hu}^2\sum_{i,j} (hu_{i,j}- hu^{obs}_{i,j})^2 + w_{hv}^2\sum_{i,j}(hv_{i,j}- hv^{obs}_{i,j})^2+w_h^2\sum_{i,j} (h_{i,j}- h^{obs}_{i,j})^2. \label{xi}
\eeqr
In this expression, we emphasize implicit  dependence of $\xi$  on time and on the set of the control parameters $p$ that is composed of 
\begin{itemize}
\item the set of initial conditions of the model $ \phi_0=\{hu\mid_{t=0},\; hv\mid_{t=0},\; h\mid_{t=0}\}$,
\item  the set of the coefficients $\alpha$ that controls the discretizations of operators near the boundary,
\item the bottom topography $H(x,y)$
\item four scalar parameters $\sigma, \mu, g, \tau_0$
\item and the Coriolis parameter $f(y)$.
\end{itemize}
Taking into account the results obtained in \cite{sw-lin}, we define the cost function as 
\beq
\costfun(p)=\int\limits_0^T  t \xi^2(\phi(p,t)) dt \label{costfn}
\eeq
that gives  bigger  importance to the difference $\xi^2$ at the end of assimilation interval.  

It should be noted here, that this cost function can  only  be used in the case of  assimilation of a perfect artificially generated data. When we assimilate some kind of real data that contains errors of measurements and is defined on a different grid, we should add some regularization term to the cost function (like the distance from the initial guess) and use some more appropriate norm instead of the Euclidean one (see, for example  \cite{UnifNotat} for details).  

The $n$th component of the gradient of the cost function can be calculated as the Gateaux derivative of an implicit function:
 \beqr
 (\nabla\costfun)_n&=&\der{\costfun}{p_n}=\int\limits_0^T t\biggl(  \der{\xi^2}{p_n}\biggr) dt=\int\limits_0^T t\biggl(  \sum_k\der{\xi^2}{\phi_k}\der{\phi_k}{p_n}\biggr) dt =\nonumber \\
 &=&2\int\limits_0^T t\biggl(  \sum_k (\phi_k-\phi^{obs}_k)  \der{\phi_k}{p_n}\biggr) dt\label{nabj}
\eeqr 
because the  derivative $\der{\xi^2}{\phi_k}$ can easily be calculated from \rf{xiphi}: $\der{\xi^2}{\phi_k}=2 (\phi_k-\phi^{obs}_k) $. The second term in \rf{nabj},  $\der{\phi_k}{p_n}$, represents the matrix of the tangent linear model that relates the perturbation of the parameter $p_n$ and the perturbation of $k$th component of the model state vector $\phi_k$. This relationship, of course, is assumed in the linear approach, that means it is only valid for infinitesimal perturbations. 

 In the classical case, when initial conditions are considered as the  only control variable, the derivative  $\der{\phi(t)}{p}=\der{\phi(t)}{\phi_0}$  is the classical tangent model that describes the temporal evolution of a small error in the initial model state. The matrix is a square matrix that is widely studied in numerous sensitivity analyses. Its singular values at infinite time limit are related to well known Lyapunov exponents that determine the model behavior (chaotic or regular) and the dimension of it's attractor.
 
 In our case,  the matrix $\der{\phi(t)}{p}$ is rectangular. It describes the evolution of an infinitesimal error in any parameter (including initial state). However, we can study its properties   in a similar way as we do with the classical tangent linear model. Its structure and composition is described in \cite{sw-lin} for the case of using coefficients $\alpha$ as control parameters and in \cite{assimtopo} for the case when the bottom topography is used to control the model's solution. 
 
The product  $\sum_k  (\phi_k-\phi^{obs}_k) \der{\phi_k}{p_n}$ in \rf{nabj} represents an unusual  vector-matrix product. To calculate this product directly we would  have to evaluate all the elements of the matrix. This would require as many tangent model runs as the size of the state vector is. So, instead of the tangent model, we shall use the adjoint one that allows us to get the result by one run of the model. Backward in time adjoint model  integration that starts from $(\phi-\phi^{obs}) $  provides immediately the product  $\biggl( \der{\phi}{p}\biggr)^*(\phi-\phi^{obs}) $ which is exactly equal to  $  (\phi-\phi^{obs}) \der{\phi}{p}$ in \rf{nabj}.  

  Using these notations, we write
\beq
\nabla\costfun= 2 \int\limits_0^T t \biggl( \der{\phi(t)}{p}\biggr)^*(\phi( p,t)-\phi^{obs}(t)) dt  \label{grad}
\eeq
where the expression in the integral is the result of the adjoint model run from $t$ to 0 starting from the vector $ (\phi( p,t)-\phi^{obs}(t))  $.

Tangent and adjoint models have been automatically generated by the Tapenade software \cite{Hascoet04},\cite{Tber07} developed by the TROPICS team in INRIA. This software analyzes the source code of the nonlinear model and produces codes of its derivative $ \der{\phi}{p}$ and of the adjoint $\biggl( \der{\phi}{p}\biggr)^*$. 

This gradient is used in the minimization procedure that is implemented in order  to find the minimum of the cost function:
\beq
\costfun(\bar p) = \min\limits_p \costfun(p) \label{min}
\eeq
Coefficients $\bar p$  are considered as coefficients achieving an  optimal parameter for the model. 
As it has been already noted, the set of parameters $p$ is composed of   the set of initial conditions of the model $ \phi_0$,
  the set of the coefficients $\alpha$ that controls the discretization of operators near the boundary,
the bottom topography $H(x,y)$
 four scalar parameters $\sigma, \mu, g, \tau_0$
 and the Coriolis parameter $f(y)$. We shall minimize the cost function  controlling as the  total set of available parameters $p$ and any possible subset comparing the efficiency of the minimization.

We use the  minimization procedure  developed by Jean Charles Gilbert and  Claude Lemarechal, INRIA \cite{lemarechal}.  The procedure uses the limited memory quasi-Newton method.

\subsection{Sensitivity estimates}

In addition to the data assimilation, we perform also the sensitivity study of the model solution to parameters enumerated in the previous subsection. We are looking for a perturbation in the model's parameters $\delta p$  that, for a given small  norm,  maximizes the norm of the perturbation of the solution at time $t$.  
\beq
\lambda(t)=\max\limits_{\delta p} \fr{\norme{\delta\phi(t)}}{\norme{\delta p}}
\eeq
We cannote that we already have  all the necessary software to estimate $\lambda(t)$. Tangent linear model $\biggl( \der{\phi(t)}{p}\biggr)$ allows us to calculate $\delta\phi(t)= \biggl( \der{\phi(t)}{p}\biggr)\delta p$. Using the scalar product that corresponds to the  norm  in the definition of the distance $\xi$ \rf{xi}, we can write 
\beqr
\lambda(t)&=&\max 
\fr{\spm{\delta\phi(t)}{\delta\phi(t)}}{\spm{\delta p}{\delta p}}= 
\max \fr{\spm{\biggl( \der{\phi(t)}{p}\biggr)\delta p}{\biggl( \der{\phi(t)}{p}\biggr)\delta p}}{\spm{\delta p}{\delta p}}= \nonumber \\
&=&\max \fr{\spm{\biggl( \der{\phi(t)}{p}\biggr)^*\biggl( \der{\phi(t)}{p}\biggr)\delta p}{\delta p}}{\spm{\delta p}{\delta p}} \label{relritz}
\eeqr
This expression is a well known  Rayleigh-Ritz ratio which is equal to the largest eigenvalue of the problem
\beq
\biggl( \der{\phi(t)}{p}\biggr)^*\biggl( \der{\phi(t)}{p}\biggr) \vartheta= \lambda(t)\vartheta \label{eigval}
\eeq
So far, we need just the maximal eigenvalue and the matrix of the problem is a self-adjoint positive definite matrix, we can solve the problem \rf{eigval} by the power method performing successive iterations 
$$
\vartheta_{n+1}=\fr{ \biggl( \der{\phi(t)}{p}\biggr)^*\biggl( \der{\phi(t)}{p}\biggr) \vartheta_n}{\norme{ \biggl( \der{\phi(t)}{p}\biggr)^*\biggl( \der{\phi(t)}{p}\biggr) \vartheta_n}},\;\; \vartheta_0=\mbox{random vector} 
$$
In the limit,  the denominator of the right-hand-side tends to the largest eigenvalue and  $\vartheta_{n}$, to the corresponding  eigenvector of the matrix. 
The principal advantage of this method consists in the fact that we do not need to calculate the matrix itself, we just need  a matrix-vector  product. So far, we have both codes for the  tangent and adjoint models; we can successively run these models and get the left-hand side of \rf{eigval}. 

We should note here that when the initial conditions of the model are used as the control parameters (i.e. $\delta p=\delta\phi(0)$), the sensitivity characteristics $\lambda(t)$ are all close to one when $t \longrightarrow 0$. It is evident because the perturbation has no time to be transformed by the model's dynamics and we get $\delta\phi(t)\mid_{t \longrightarrow 0}=\delta\phi(0)=\delta p$. 

When any other model parameter is used as the control and the error growing time is small, all   $\lambda(t)$ are vanishing. This is also clear: the model's dynamics has no time to transmit the perturbations from the parameters to the solution. The perturbation of the solution remains, consequently, close to zero as well as the value of  $\lambda(t)\mid_{t \longrightarrow 0}=0$.  

In order to make   the behavior of the sensitivity characteristics  uniform with different parameters, we shall use  $\lambda(t)-1$ every time when the initial model's state is considered as the control parameter.  

Another point that we should emphasize, concerns the physical dimensions of parameters. So far we want to compare the sensibility of the model to various parameters; we should be careful with bringing them to the dimensionless form because the choice of the characteristic values influences the result. In this paper, we choose to measure all the perturbations in fractions of the original non-perturbed parameter. This will ensure relatively uniform  weighting of perturbations. That means the perturbation of the Coriolis parameter is normalized by $f_0$, the value of this parameter in the middle of the basin; the perturbation of the bottom topography is normalized by $H_0$, the average depth; perturbations $\delta\mu,\; \delta\sigma,\; \delta g$ and $\delta\tau_0$ are respectively normalized by   $\mu,\; \sigma,\;  g$ and $\tau_0$. As it has been already noted, perturbations of the initial conditions $\delta\phi_0$ and of  the model state vector $\delta\phi(t)$ are already dimensionless, being obtained form the model's variables using  weights  $w_{hu}=w_{hv}=\fr{1}{H_0\sqrt{gH_0}}$ and  $w_h=\fr{1}{H_0} $.  Coefficients $\alpha$ that are used to control the discretization of operators are also already dimensionless, having characteristic values around one (+1 or -1 in the second order derivatives and $1/2$ in the interpolations) they are used without normalization. 

Thus, the Rayleigh-Ritz ratio \rf{relritz} and the eigenvalues of problem \rf{eigval} become dimensionless, but they keep the dependence on the normalization constants. 

\section{Model in a square box. }

We start from the data assimilation in frames of the very well studied "academic" configuration. 
Several experiments have been performed with the  model  in a square box of side length $L=2000$ km driven by a steady, zonal  wind forcing with a classical sinusoidal profile
$$
\tau_x=\tau_0 \cos \fr{2\pi (y-L/2)}{L}
$$
that leads to the formation of a double gyre circulation \cite{LPV}. The attractor of the model and the bifurcation diagram  in a similar configuration has been described in \cite{simmonet2}. Following their results, we intentionally chose the model's parameters to ensure chaotic behavior. The maximal wind tension on the surface is taken to be $\tau_0=0.5\fr{dyne}{cm^2}$.  The coefficient of Eckman dissipation and the  lateral friction coefficient  are chosen as
$\sigma=5\tm 10^{-8}s^{-1}$ and  $\mu=200\fr{m^2}{s}$ respectively. 

As it has been already noted, the Coriolis parameter is a linear function in $y$ with  $f_0=7\tm 10^{-5}s^{-1}$ and $\beta=2\tm 10^{-11} (ms)^{-1}$. The reduced gravity and the depth are respectively equal to $g=0.02\fr{m}{s^2},\;H_0=1000m$.

The resolution of the model in this section is intentionally chosen to be too coarse to resolve the Munk layer \cite{Munk}  that is characterized by the local equilibrium between the $\beta$-effect and the lateral dissipation. Its characteristic width is determined by the Munk parameter $ d=2\biggl(\fr{\mu}{\beta}\biggr)^{1/3}$ which is equal to 42 km in the present case.
The model's grid is composed of 30 nodes in each direction, that means the grid-step is equal to 67 km, that is more than  the Munk parameter.  Thus, there is only one grid node in the layer and the solution exhibits spurious oscillations near the western boundary due to unresolved boundary layer.  

Artificial ``observational`` data are generated by
the same model with all the same parameters but with 9 times finer resolution  (7.6 km  grid step). The  fine resolution model, having 7 nodes in the Munk layer, resolves  explicitly the layer and must have no spurious oscillations. All nodes of the coarse grid belong to the fine grid, consequently, we do not need to interpolate ''observational'' data to the coarse grid. We just take values in  nodes of the high resolution grid that correspond to nodes on the coarse grid. 

The model on the fine grid has been spun up from the rest state during 3 years. The end of spin up was used as the initial state for the further integration of the model. From the result of this integration we have extracted values of all three variables at all grid points that belong to the coarse grid (as it has been noted, the grids have been chosen so, that all grid points of the coarse grid belong to the fine grid). This set   is used as artificial observations in the following experiments. 

So far the model is nonlinear with intrinsically instable solution, there is no hope to obtain close solutions in  long time model runs because any difference (even infinitesimal) between two models grows exponentially in time. Consequently, we have to confine our study to the analysis of a short time evolution of the model's solution simulating the forecasting properties of the model.  

All operators in the model are approximated either with the second or with the fourth order accuracy in the interior of the domain by \rf{intrnlsch} or \rf{intrnlsch4}.  Initial guess for the coefficients $\alpha$ is  defined to satisfy the second order scheme. That means the expression \rf{bndsch}, that is used to interpolate functions and to calculate their derivatives near the boundary, is written with $ \alpha_0=0$. Coefficients $\alpha$ are defined as $\alpha^{D}_{1}=-1,\; \alpha^{D}_{2}=1$ for all derivative operators and $\alpha^{S}_{1}=\alpha^{S}_{2}=1/2$ for all interpolations. That gives, for example, the value of the derivative of $u$ at the point $i=1/2$ as $(D_x u)_{1/2,j-1/2}=\fr{u_{1,j-1/2}-u_{0,j-1/2}}{\delta x}   $.

\subsection{Data assimilation}

Final values of the cost function obtained in experiments of the assimilation of  artificially generated data into the shallow water model are shown in the table 1. All the experiments are carried out in the same conditions with the same data. The assimilation window has been chosen as 5 days interval.  This time interval corresponds well to the characteristic time of the model's physics. Gravity waves, with their velocity equal to $\sqrt{gH_0}=4.5 \fr{m}{s}$, cross the 2000 km box in 5 days.

  As the initial guess for the initial conditions we use the state vector of the high resolution model reduced on the coarse grid. This state is also used as the initial conditions in all other   assimilation experiments with other control parameters.  Noted above values of the model's parameters (flat bottom topography, linear in $y$ Coriolis parameter and  scalar parameters ($\mu, \; \sigma, \; \tau_0,\; g$) are used as the initial guess in the experiments that control these parameter; otherwise, we simply use these parameters in the model. 

As it has been noted, we control the parameters in the second and the fourth order model. With no assimilation at all, the solution of  the fourth order model is "worse" than the solution of the second order one. That means the model's trajectory moves away from observations more rapidly and produces bigger cost function value.  The reason of this is clear: high order scheme works worse when principal physical scales are not resolved explicitly. Indeed,  the coarse grid of the model does not resolve the Munk boundary layer. The grid step $\delta x=67$ km is bigger than  the Munk parameter $d=2(\mu/\beta)^{1/3}=42$ km. But, it is the ratio $\biggl(\fr{\delta x}{d}\biggr)^{n}$ where $n$ is the order of approximation  that determines the approximation error. Higher  the order of approximation, bigger is this ratio and bigger is the error in the approximation of the boundary layer. Consequently, it is more important to identify an optimal numerical scheme for approximation of the boundary layer for the fourth order model. 
  Indeed, if we assimilate data and find optimal parameters of the model, the fourth order model becomes comparable and even "better" (allowing lower cost function's value)  than    the second order one. 
  
  However, the influence of different parameters on the solution is not the same. 
Comparing the cost function's values at the end of minimization procedure, we can say that the model is the most flexible with respect to the control of initial and boundary conditions. The cost function value can be divided by 3 or even 4 in the minimization procedure.  Controlling all other parameters, we cannot achieve such a low cost function. We must note also that  the  control of coefficients $\alpha$ is much more expensive than the control of initial conditions:  5 and even 10  times more iterations are required for the minimization to converge. 

Obviously, the best result is  obtained in the joint control of all available parameters of the model. The cost function value is divided by 10 but the number of iterations exceeds 100.  

\begin{table}
\begin{tabular}{|l|cc|cc|}
\hline
 &\multicolumn{2}{|c|}{Second order interior}&\multicolumn{2}{|c|}{Fourth order interior}\\
Control &$\costfun_{final}$&Iterations&$\costfun_{final}$&Iterations\\
\hline
Nothing                    & 2.51 & ---&3.29& ---\\
Initial state $\phi_0$    & 1.04 & 14 & 0.96 & 13\\
Topography $H(x,y)$       & 1.81 & 5  & 2.37 & 7\\
 Coriolis $f(y)$          & 2.15 & 57 & 2.81 & 54\\
 Scalars $\sigma,\mu,\tau_0,g$& 2.20 & 33 & 3.06 & 31\\
Boundary $\alpha$         & 1.41 & 113& 0.95 & 124\\
{\small All parameters together}  & 0.34 & 115& 0.28 & 149\\
\hline
\end{tabular}
\label{tab1}
\caption{Number of iterations and final values of the cost function in the data assimilation experiments with different control parameters}
\end{table}

Along with the value of the cost function in the assimilation window, we examine the behavior of the model beyond the window. In fact, assimilating known data we can find optimal model parameters, but the optimality is guaranteed in the assimilation window only. However, if the model is developed to provide a forecast, it has to use the observational data in the past and predict the future. In the described test case,  consequently, it is more interesting to see the behavior of the solution after the end of assimilation, i.e. beyond the assimilation window. 

In order to see whether optimal parameters can improve the model's  behavior  even when the assimilation is over, we examine the difference $\xi(t)$  between the solution and observations over 20 days interval, i.e. during 15 days next to the assimilation window. The evolution of the  difference $\xi(t)$ for different optimal subsets of the controllable set is shown in \rfg{evol-xi.sq}. Upper solid line in this figure represents the distance "observations-model" with no assimilation at all. Default parameters, described above,  are used in this model run.  

\begin{figure}[h]
  \begin{center}
  \begin{minipage}[l]{0.45\textwidth} 
  \centerline{\includegraphics[angle=0,width=0.95\textwidth]{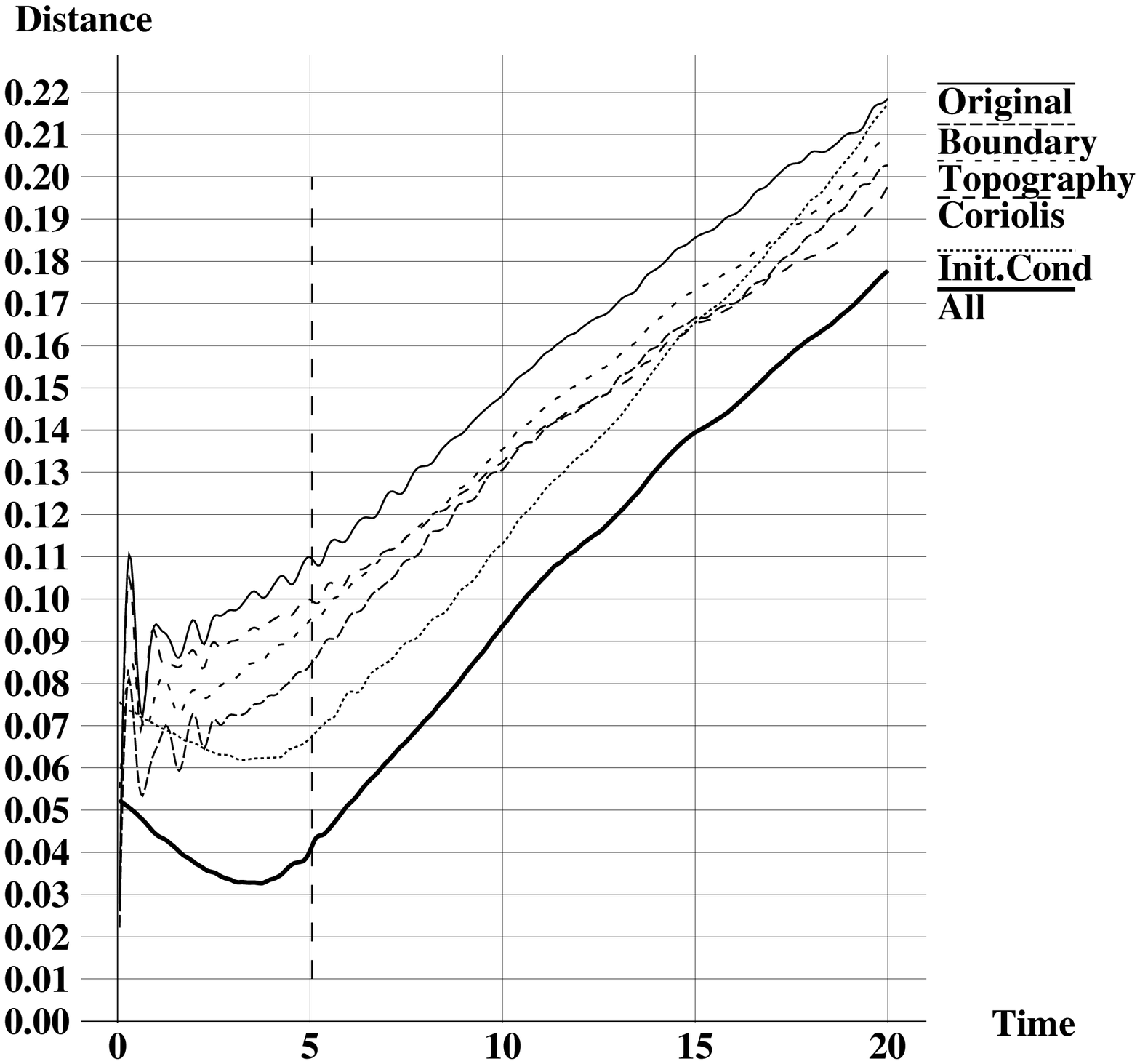}}
  \end{minipage} 
    \begin{minipage}[r]{0.45\textwidth} 
  \centerline{\includegraphics[angle=0,width=0.95\textwidth]{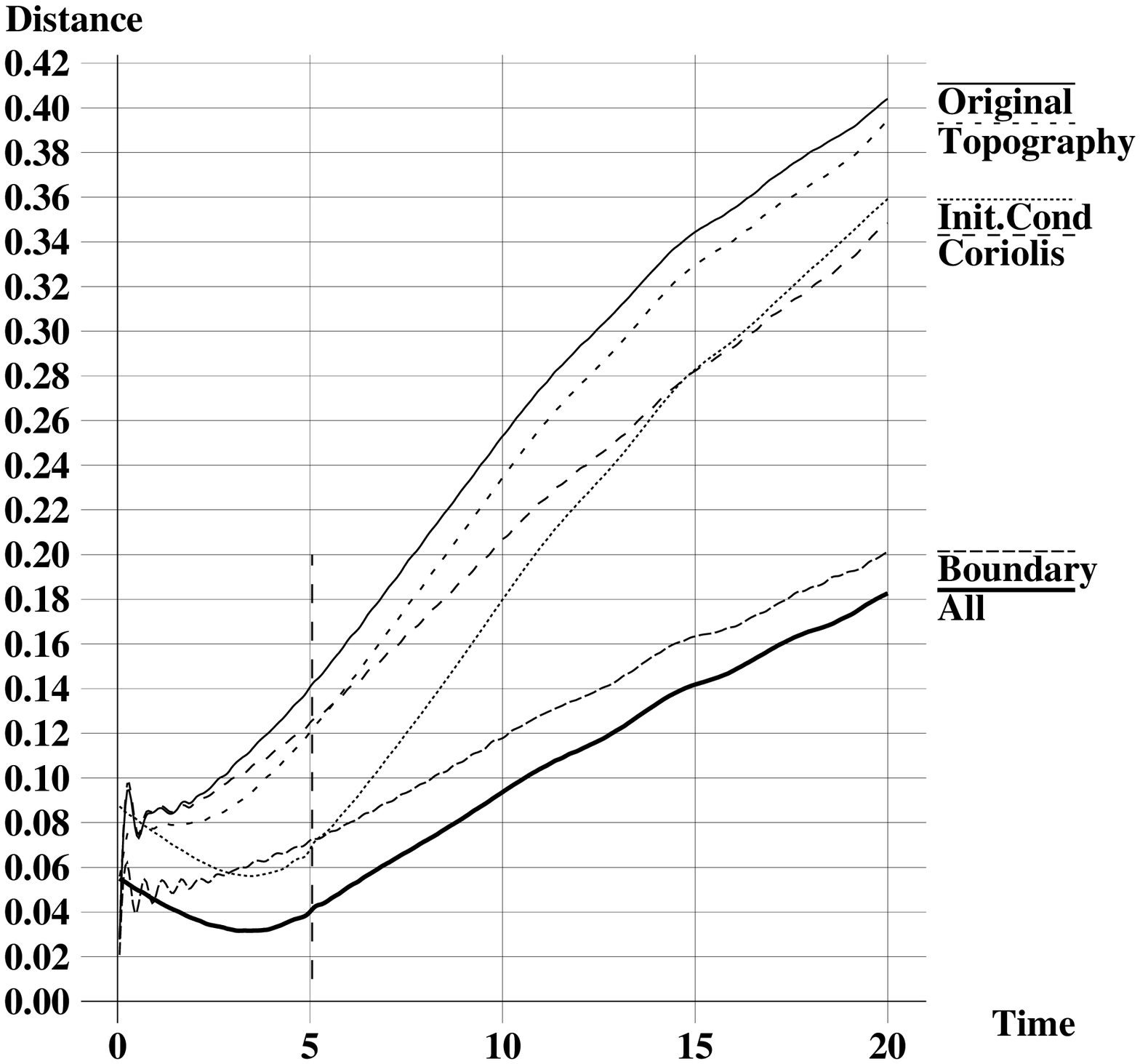}}
  \end{minipage} 
    \caption{ Evolution of the distance $\xi(t)$   \rf{xi} during  and after assimilation obtained with the second order model (left) and the fourth order model (right). }
  \end{center} 
\refstepcounter{fig}
\label{evol-xi.sq}
\end{figure}

Analysing the figure  \rfg{evol-xi.sq}, we cannote several differences in the influence of model parameters on the solution. As it has been already seen,  the flexibility of the model is small with respect to  the bottom topography, scalar coefficients and the Coriolis parameter (joint control of these two parameters is presented). As a consequence, in the assimilation window ($0\leq t \leq 5$ days) corresponding lines are close to the original line obtained with no assimilation at all. 
Beyond the window, these lines remain parallel to the original line keeping the difference that has been obtained in the minimization. In some cases,  the  increase of the distance from observations can even be less rapid than the increase of the solid line. This fact indicates that the optimal values of parameters obtained in the assimilation window remain to be optimal even beyond the window allowing us to use them for the forecast. 

On the other hand, the model started from the optimal initial state $\phi_0$ exhibits a very different behavior in the window ensuring low cost function value. However, just after  assimilation end,  the distance from observations starts to increase and moves toward the solid line obtained with the default parameters. That means, controlling the initial state of the model allows us to improve the model in the assimilation window and the  short range forecast  (5-10 days in this experiment) as well,  but has almost no influence on  longer forecasts (15 days here). 

The most spectacular result of the model improvement can be obtained in the experiment that controls coefficients $\alpha$ for the fourth order model. The value of the distance from observations is divided by 2 at $T=20$ days. That means optimal parametrization of boundary conditions remains optimal after the end of assimilations and can help to improve the model bringing the solution closer to the solution of a high resolution model used to generate artificial observations in this experiment.  This can be explained by the improved accuracy of the fourth order approximation in the interior of the domain accompanied by optimal boundary conditions that are really necessary for this model. 

Thus, we see in this experiment that if the model's dynamics suffers from low resolution and other numerical errors,  better forecast is achieved by controlling the model's operators rather than initial conditions.

\subsection{Sensitivity estimates.}
The third way of the sensitivity analysis consists in solving of the eigenvalue problem \rf{eigval} and analyzing  $\lambda(t)$ on different  scales of error growing time from about $10$ minutes  ($10^{-3}$ day) to more than one year (500 days). Characteristic time scale (5 days during which the gravity waves cross the domain) is situated in the middle of this interval.  Performed experiments with the second and the fourth order models show that $\lambda(t)$ are very close to each other. So, we plot only one of them in \rfg{lambda.sq}, and namely sensitivity estimates of the second order model. 
As it has been already noted, $\lambda(t)-1$ is plotted in the case when initial conditions are considered as the parameter.

\begin{figure}[h]
  \begin{center}
  \begin{minipage}[r]{0.46\textwidth} 
  \centerline{\includegraphics[angle=0,width=0.95\textwidth]{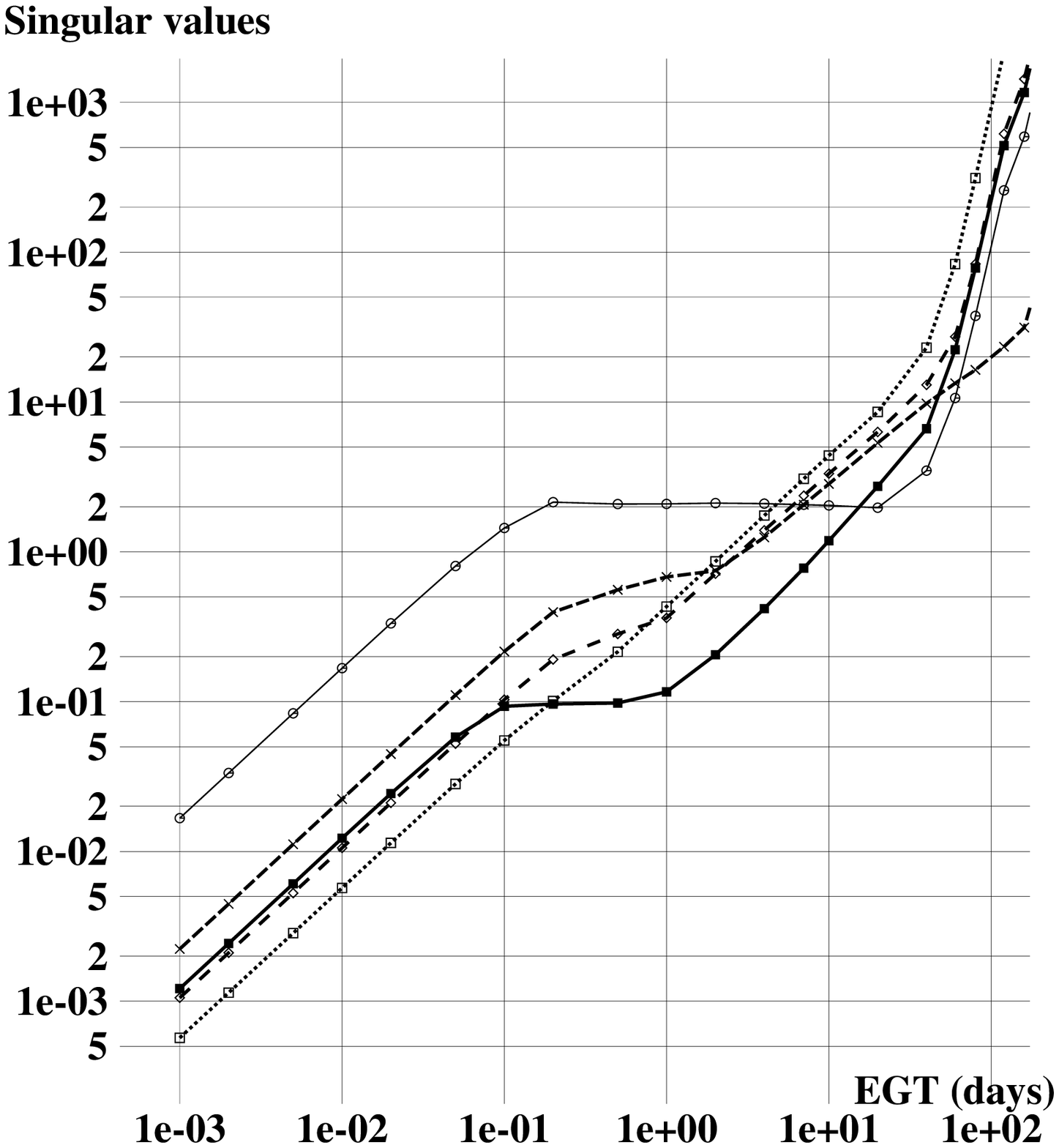}}
  \end{minipage} 
  \begin{minipage}[r]{0.52\textwidth} 
  \centerline{\includegraphics[angle=0,width=0.95\textwidth]{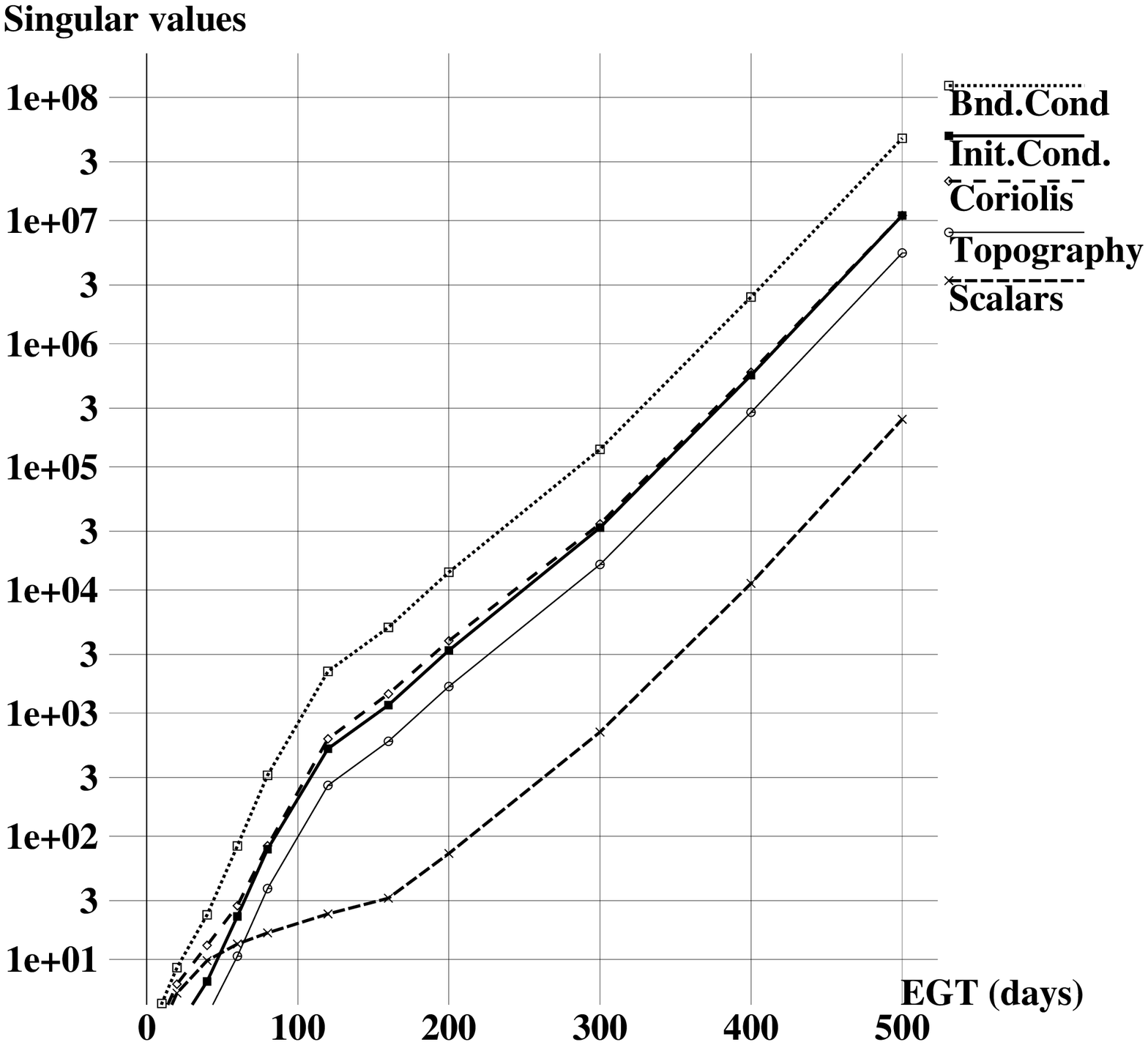}}
  \end{minipage} 
  \end{center} 
\refstepcounter{fig}
  \caption{Sensitivity characteristics  $\lambda(t)$  as functions of the error growing time (Log-Log coordinates (left) and Log-Linear coordinates (right).    }
\label{lambda.sq}
\end{figure}

Analyzing the figure \rfg{lambda.sq}, we can see that  three  time scales can be clearly distinguished  for the sensitivity characteristics of the model. The first, short time scales, approximately from 0  up to  2-3 hours is characterized by the  linear growth of   $\lambda(t)$.  We should note that the growth is not only linear in the logarithmic coordinates, but the slope is equal to 1. Consequently, $\lambda(t)$ is proportional to $t$.  Indeed, the model behaves as a linear model on these scales,  the model's solution can be well approximated by just one step of the numerical time scheme. 

The second time scale  that can be distinguished in the  figure \rfg{lambda.sq} corresponds to error growing times from 2-3 hours to 10-100 days. On these time scales we see slower growth of the sensitivity characteristics $\lambda(t)$ and, sometimes, no growth at all. These time scales are characterized by the modification of the stable-instable subspaces of the model. Instable space on short time scale is not the same as for long time evolution. Short time instabilities are usually localized in space, while long time eigenvectors of \rf{eigval}  possesses a global structure.   On the medium time scales we see the transformation of the instable space of the quasi-linear model to the instable space of the non-linear chaotic model. This modification stipulates the slowdown (and even stagnation) of growth of $\lambda(t)$.

The third time scale corresponds to the error growing times more than 100 days. On these scales the model exhibits non-linear chaotic behavior with exponential growth of all $\lambda(t)$. In order to zoom these time scales, we plot the same data in the Log-Linear coordinates in \rfg{lambda.sq}  on the right. One can see that the growth on this time scale is purely exponential with the same exponent $\lambda(t)=A\exp(0.027t)$. The multiplier $A$ is particular for each parameter, but the exponent is always the same.  This confirms the remark made  in \cite{assimtopo}, \cite{sw-lin}: no matter how the perturbation was introduced into the model, its long-time growth  is determined by the model's dynamics.

Comparing the evolution of the sensitivity of the model to different parameters, we see that on small scales it is the bottom topography that the model is the most sensitive to (thin solid line in  \rfg{lambda.sq}). An error in the topography produces 13 times bigger perturbation in the model state than a similar error in the model's initial conditions (thick solid line in  \rfg{lambda.sq}). However, $\lambda(t)$ does not grow at all on medium scales due to significant changes in the eigenvector's pattern. This leads to  the fact that on long scales, the sensitivity of the model to the bottom topography is about 2 times lower than the sensitivity to initial conditions. 

On the other hand, the sensitivity of the model to the discretization of operators near the  boundary exhibits the opposite behavior. On short scales, corresponding $\lambda$ is 2 times lower than    $\lambda$ obtained for perturbations of  $\phi_0$, but there is no stagnation of the growth on the middle scales. As a result, we see that the model is 4 times more sensitive to $\alpha$ than to $\phi_0$ for long error growing times.

\section{Model of the Black Sea. }

In this section we use the same model, but all the parameters are defined to describe the upper layer circulation of  the Black sea. Configuration of the model  and  observational data have been kindly provided by Gennady Korotaev from the  Marine Hydrophysical Institute, National Academy of Sciences of
Ukraine, Sevastopol, Ukraine. This configuration is described in \cite{korot-model}. 

The model grid counts $141\tm 88$ nodes  that corresponds to the  grid box of dimension 7860 m and 6950 m in $x$ and $y$ directions, respectively. A  15 minutes  time step is
used for integration of the model.  The Coriolis parameter is equal to  $f_0= 10^{-4}s^{-1}$ and $\beta=2\tm 10^{-11} (ms)^{-1}$. Horizontal viscosity is taken as $\mu=50 m^2 s^{-1}$. Using a typical density difference between upper and underlying layers of  $3.1 kg/m^3$ , and unperturbed layer thickness of $H_0 = 150 m$, the Rossby radius of deformation is estimated
at about 22 km and the reduced gravity value $g=0.031 m/s^2$. The grid therefore resolves the mesoscale processes reasonably well.

The model has been forced by the ECMWF wind stress data, available as daily averages for the years 1988 through
1999. 
Dynamical sea level reconstructed in \cite{korot-altim} was used as  observational data in this section. These data   have been collected in ERS-1 and TOPEX/Poseidon missions and preprocessed by the NASA Ocean Altimeter Pathfinder Project, Goddard Space Flight Center. Observational data are available from the 1st May 1992 until 1999. These data have been linearly interpolated to the model grid.

So far  the sea surface elevation is the only observational variable available in this experiment, we put $w_{hu}=w_{hv}=0$ in \rf{xi}. Consequently, the difference between the model's solution and observations is calculated taking into account the variable $h$  only.  

As it has been already noted, absence  of  observational data for the velocity fields brings   us to modify the cost function. We have to add the background term in the cost function in order to  require the velocity field to be sufficiently smooth. Otherwise, lack of information about velocity components in observational data would result in a spuriously irregular field obtained in assimilation. To ensure necessary regularity of $hu$ and $hv$ we add the distance from the initial guess to the cost function \rf{costfn}. In order to emphasize the requirement of smoothness,  this distance is measured as an enstrophy of the difference between the initial guess and the current state: 
\beq
\costfun_{smooth}=  \sum_{i,j}\biggl(\der{(hv_{i,j}-hv_{i,j}^0)}{x}-\der{(hu_{i,j}-hu_{i,j}^0)}{y}\biggr)^2  \label{costfn-smooth}
\eeq
where $hu^0,\; hv^0$ denote  flux components of the initial guess of the minimization procedure. 

Moreover, using real observational data requires to add at least one another term to the cost function. One can see that in 
the Figure 2 in \cite{korot-altim},  spatially averaged  sea surface elevation of the Black sea exhibits a well distinguished seasonal cycle. That means the mass is not constant during a year, it decreases in autumn and increases in spring. Consequently, if we assimilate data during a short time (a season or less), we assimilate also the information about the  mass flux specific for this season. This flux cannot be corrected later by the model because the discretization of operators near the boundary (that controls the mass evolution) is obtained once for all seasons.  The mass variation of the Black sea  reaches 25 centimeters of the sea surface elevation. Assimilating data  within one season may, consequently,  result in a persisting increase or decrease of the sea level of order of 50 cm  per year.   To avoid this spurious change of the total mass, we must either take the assimilation window of at least one year, or prescribe the mass conservation to the model's scheme. One year assimilation window is computationally expensive and is not justified by the model's physics. On the other hand, prescribed mass conservation removes just the sinusoidal seasonal variation, allowing us to keep all other processes and to choose any assimilation window we need. 

To correct the mass flux of the model, we add the following term to the cost function
\beq
\costfun_{mass}= \int\limits_0^T \biggl(\sum_{i,j}(h_{i,j}(t)-h_{i,j}(0))\biggr)^2 dt \label{costfn-mass}
\eeq
Similar to \rf{costfn-smooth}, this term also ensures the regularity of the solution. It can be noted here that  other terms may be added to the cost function in order to make a numerical scheme  energy and/or enstrophy conserving, but we do not use them in this paper. 

The total cost function in this section is composed of three parts: \rf{costfn}, \rf{costfn-smooth} and \rf{costfn-mass}:
\beq
\costfun_{total}= \costfun+\gamma_1\costfun_{smooth}+\gamma_2\costfun_{mass}\label{costfn-total}
\eeq
Coefficients $\gamma$ are introduced to weight the information that comes from observational data (with $\costfun$) and an a priori knowledge about mass conservation and regularity of the solution. 

This modification of the cost function results, of course, in additional terms in the gradient:
\beq
\nabla  \costfun_{total}= \nabla\costfun+ 2\gamma_1 \biggl( D_y^*D_y (hu-hu_0) + D_x^*D_x (hv-hv_0)\biggr) 
+2\gamma_2 \sum_{i,j} \biggl(\eta_{i,j}(t)-\eta_{i,j}(0)\biggr).
\eeq

The model is spun up  from the beginning of 1988 to May 1992 using the wind tension data on the surface. The state corresponding to the 1st of May 1992 12h GMT is used as the initial guess in the data assimilation procedure controlling initial conditions of the model.  The assimilation controls the initial conditions $\phi_0$ only with the assimilation window $T=1$ day and the regularization parameter $\gamma_1=0.04$.   Such a short window was chosen in order to get almost instantaneous state of the model to be used in further experiment as an initial state. 

In this paper we chose a  $T=30$ days window which is longer than synoptic time scales. The minimization of the cost function has been accompanied by the mass preserving correction \rf{costfn-mass} with $\gamma_2= 0.01$. 

\subsection{Data Assimilation}

Like in the previous section,  we examine the influence of the model parameters from two points of view. First, we assimilate observational data  observing  the cost function value in the assimilation window.  This experiment reveals the flexibility of the model with respect to a parameter; it shows how close  the solution can be with the optimal values of this parameter. Analysis of the cost function beyond the window shows the capability of the parameter to improve the forecast quality.    And second, we apply the classical sensitivity analysis of the model solution with respect to parameters calculating the largest eigenvalue $\lambda(t)$ of  problem \rf{eigval} that shows the influence of a small error in parameter on the solution of the model at time $t$. 

As we have already noted, we  consider the same model,  but there is a principal difference with the previous case of the model in a square box. The behavior of the model  solution is not chaotic in this configuration. Variability of the model is generated directly by the variability of the wind stress on the surface.  Consequently, we can compare particular trajectories of the model on any time interval because their evolution is stable without exponential divergence. Thus, we can hope that  assimilating data in a relatively short window  allows us to bring the model's solution closer to observation for a long integration period. 

The flexibility of the model is illustrated in \rfg{blk} on the left. We perform the data assimilation experiment with 30 days assimilation window using parameters described above as initial guess.  Due to high CPU time of the data assimilation, we limit the number of iterations of the minimization procedure by 20. Thus, we have similar  and reasonable computational cost in each experiment. 

 One can see the model is the least flexible when we control the bottom topography and the most flexible with respect to the control of coefficients $\alpha$. We can get 2 times lower distance solution--observations controlling discretization of operators near the boundary than controlling the topography. 
 One cannote that  Black sea model is less flexible with respect to the initial conditions than the model in the square box. This fact is due to the additional regularization term \rf{costfn-smooth} that is added to compensate the lack of observational data for $u$ and $v$ variables. On the other hand we can see increased flexibility with respect to the boundary conditions despite the presence of the forced mass conservation \rf {costfn-mass}. This fact indicates the importance of controlling $\alpha$ for a model in a realistic domain.

\begin{figure}[h]
  \begin{center}
  \begin{minipage}[r]{0.48\textwidth} 
  \centerline{\includegraphics[angle=0,width=0.95\textwidth]{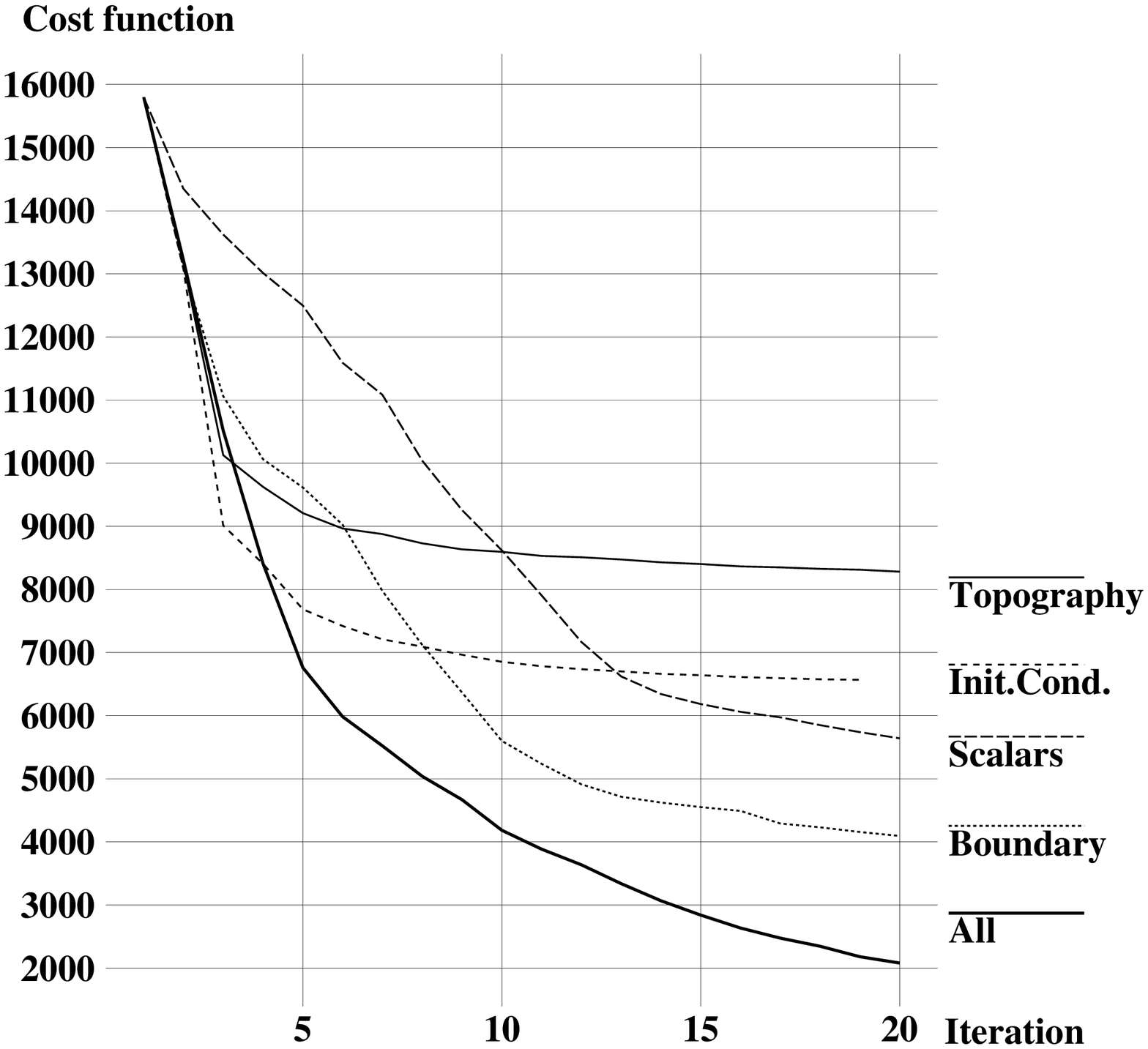}}
  \end{minipage} 
   \begin{minipage}[r]{0.48\textwidth} 
  \centerline{\includegraphics[angle=0,width=0.95\textwidth]{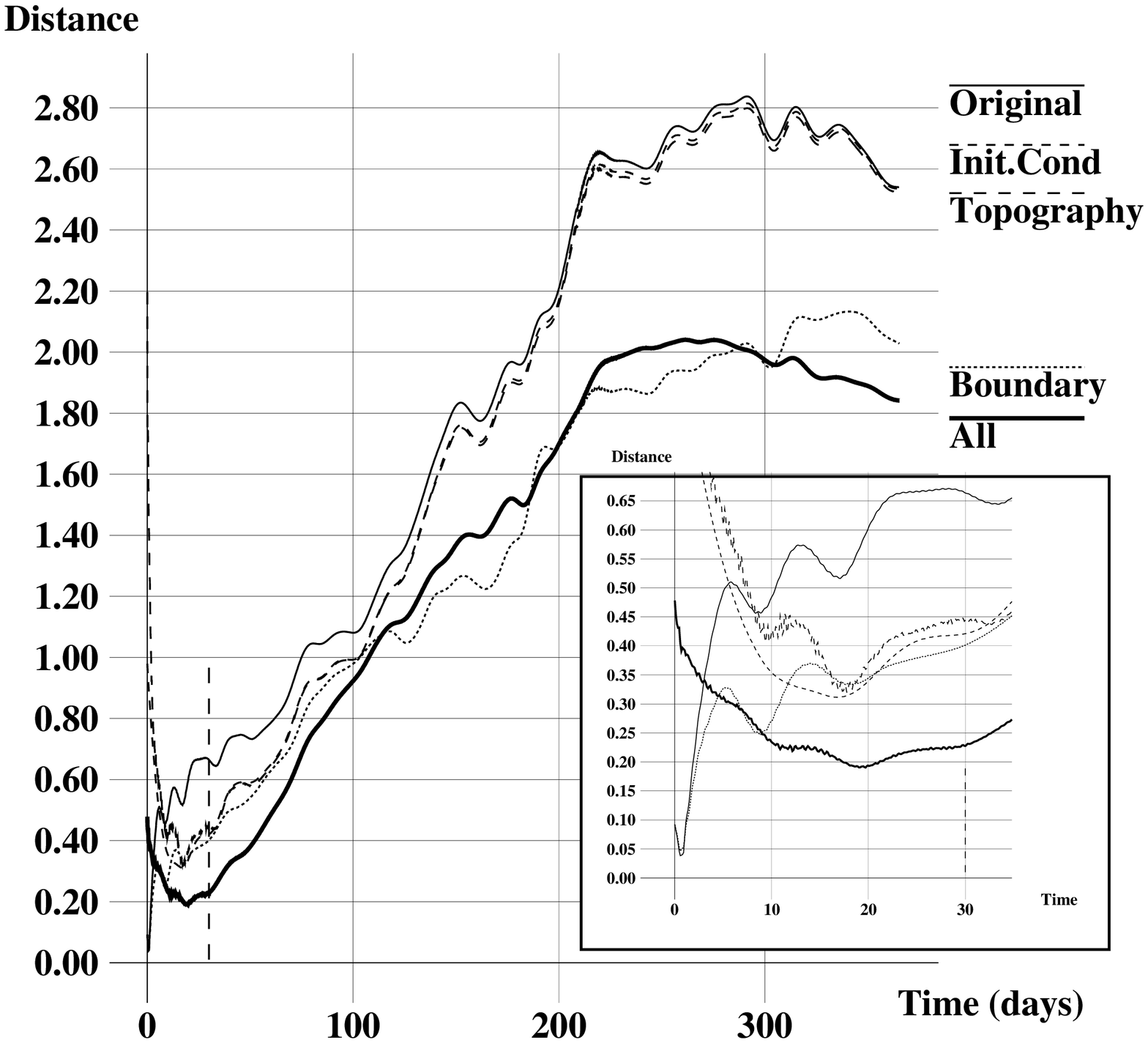}}
  \end{minipage} 
  \end{center} 
  \caption{Convergence of the cost function in the minimization procedure (left) and evolution of the distance "model-observations during 1 year (right).  }  
\refstepcounter{fig}
\label{blk}
\end{figure}

The evolution of the distance "model--observations" $\xi(t)$ is shown in the \rfg{blk} on the right. We assimilate the data in the 1 month window and integrate the model for the next 11  months.  Assimilation window is zoomed in this figure in order  to better distinguish different lines. One can see several differences with the behavior of the model in the square box. Controlling the initial state and the topography drastically  increases the distance at the origin $\xi(0)$ indicating the initial guess is far from being optimal. The behavior of the model's solution with optimal initial point, with  optimal topography and with optimal scalar parameters (not shown in the figure because the line is indistinguishable from other lines) are very similar as in the window and beyond. 

There are also common points with  the experiment in the square box.  It is the control of coefficients $\alpha$ that allows us to improve a long range forecast. Optimal initial point for the 30 days assimilation window influences the forecast during 100 days and after that there is no significant difference between the model with optimal parameters and the model with the default ones. Another common point consists in an expected  fact that the joint control of all parameters ensures the lowest distance from observations as in the window and beyond.

\subsection{Sensitivity estimates.}

And finally, we consider the sensitivity characteristics $\lambda(t)$ of the model to its parameters. The dependence of $\lambda(t)$ on the Error Growing Time (EGT) is shown in the  \rfg{lambda.blk}. The figure is also divided into  two parts:  short time scales are zoomed on the left and long time scales  on the right.   Comparing this figure with the  \rfg{lambda.sq} we see that there is no general exponential growth of $\lambda(t)$ on long time scales. Moreover, the perturbation of initial conditions decreases on long time scales and corresponding $\lambda(t)$ become smaller than one (that is  why we cannot plot $\lambda(t)-1$ in logarithmic coordinates for $t>20$).  This fact shows that  the model solution is not chaotic and the variability of the solution is only due to the variability of the wind stress on the surface. The sensitivity of the model to other parameters does not show any regular behavior on long time scales. While $\lambda(t)$  that  correspond to the boundary conditions and to the Coriolis parameter are growing with time (but not really exponentially), the sensitivity to scalar model parameters oscillates and the sensitivity to the bottom topography stagnates. 

However, comparing \rfg{lambda.blk} and \rfg{lambda.sq} one can remark also several common points. We can also distinguish  short, medium  and long  scales on which the behavior of  $\lambda(t)$ is different.  
Obviously, linear error growth  is also observed on short scales.  On these scales, the model solution is also the most sensitive to topography perturbations and the sensitivity to boundary conditions is also smaller than the sensitivity to initial conditions.  On the medium time scales   we can also see common points with the square box. Sensitivity  to the topography reaches the value $\lambda(t)=2$ and  stagnates after that; $\lambda(t)$  corresponding to the boundary conditions and the Coriolis parameter continue to grow.

\begin{figure}[h]
  \begin{center}
  \begin{minipage}[r]{0.46\textwidth} 
  \centerline{\includegraphics[angle=0,width=0.95\textwidth]{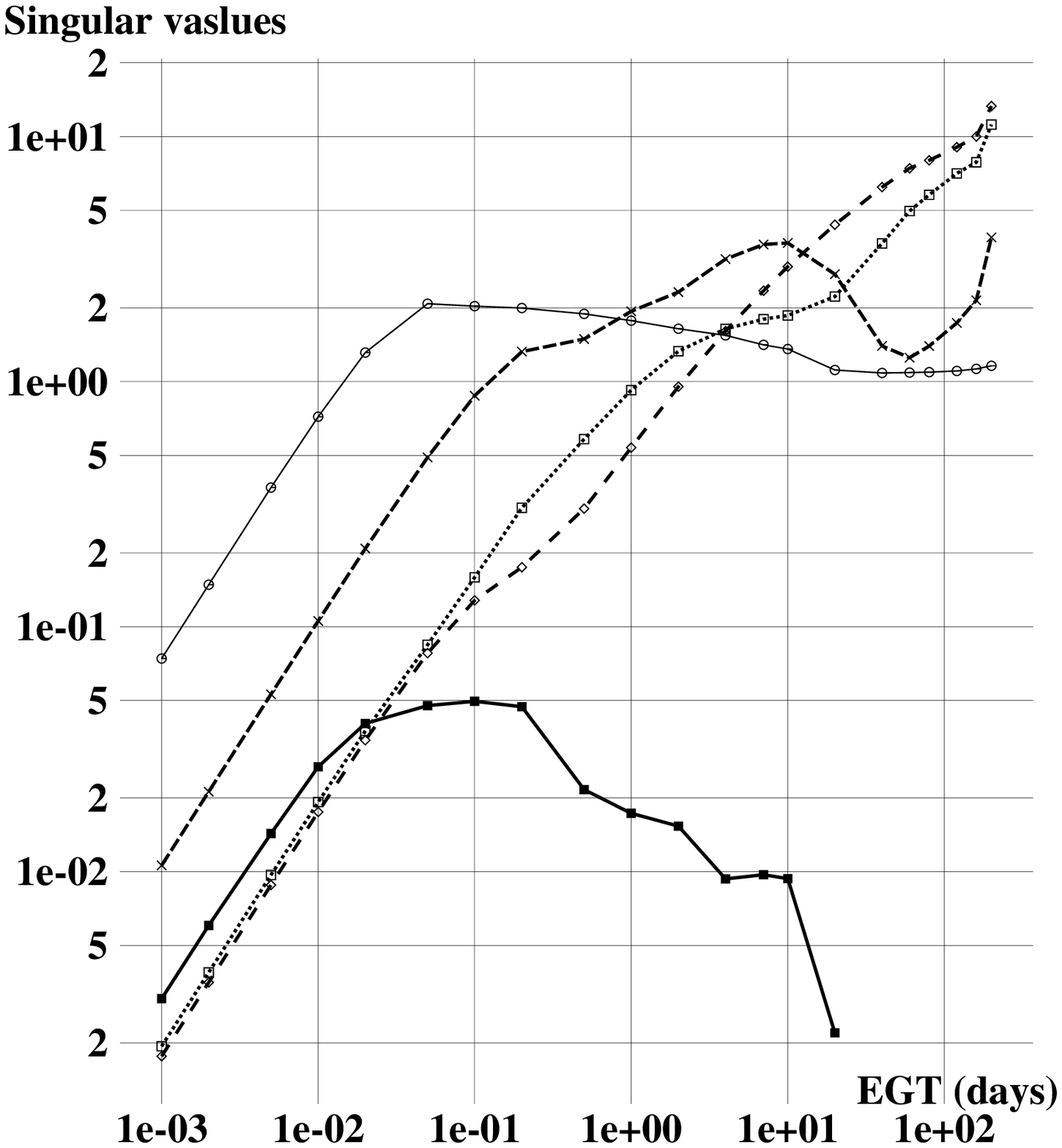}}
  \end{minipage} 
  \begin{minipage}[r]{0.52\textwidth} 
  \centerline{\includegraphics[angle=0,width=0.95\textwidth]{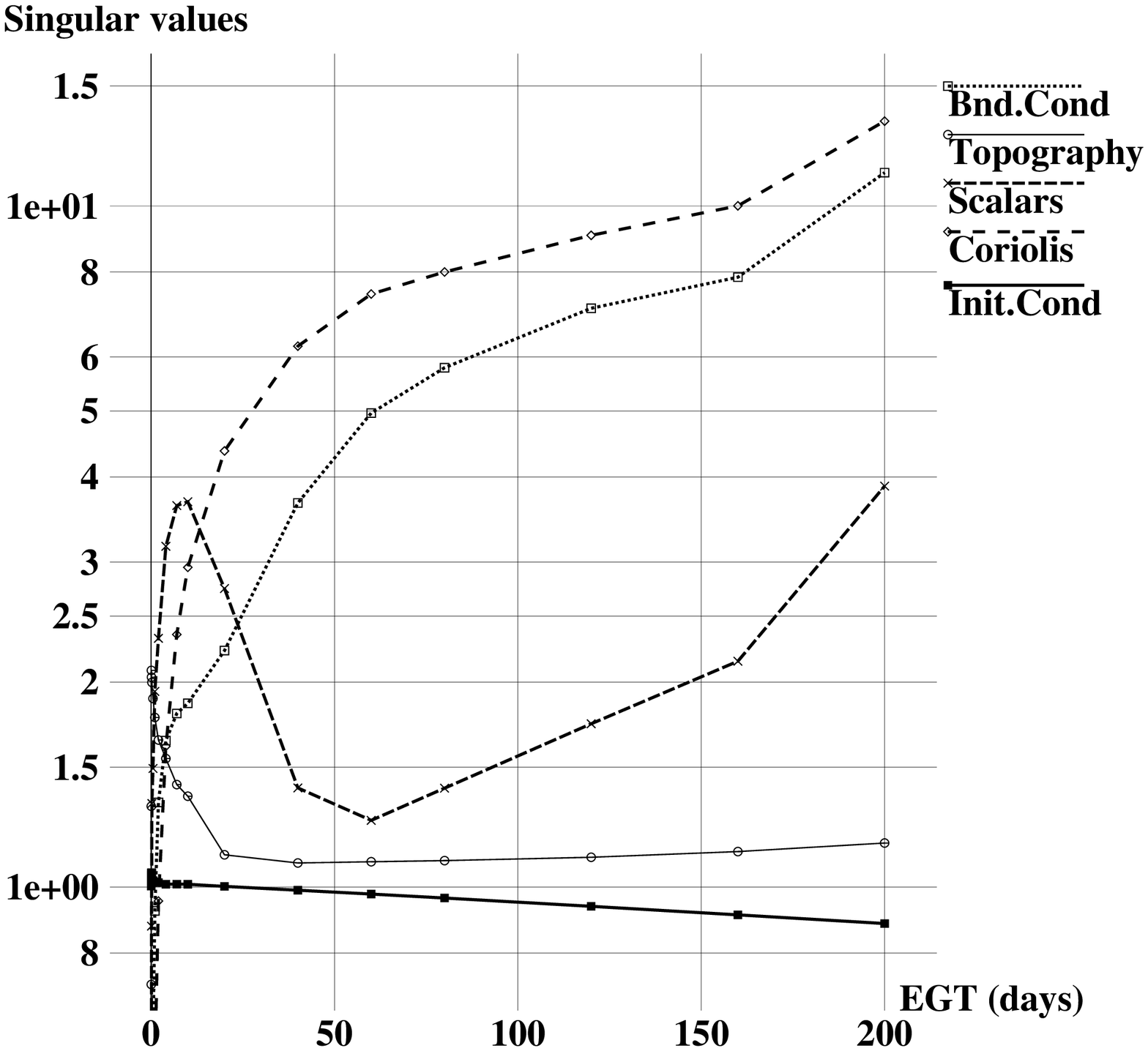}}
  \end{minipage} 
  \end{center} 
\refstepcounter{fig}
  \caption{Sensitivity characteristics  $\lambda(t)$  as functions of the Error Growing Time (Log-Log coordinates (left) and Log-Linear coordinates (right).    }
\label{lambda.blk}
\end{figure}

\section{Conclusion}

The comparative study presented in this paper shows the  influence of different model parameters on the solution. We do not show optimal parameters of the model that have been obtained in the assimilation process, nor the most sensitive patterns obtained as singular vectors of \rf{eigval} reserving all these data and analysis  to a more technical study and discussion. In this paper we do just a comparison of the sensitivity of the model to the set of its internal and external parameters. 

The study is confined to the analysis of a low resolution model with a rather limited physics.  Consequently we must acknowledge that the results may be valid only in the described case. Additional physical processes (baroclinic  dynamics,  variable density due to heat and salinity fluxes, etc.) may modify  results of this study revealing other parameters the model may be sensitive to.   

The main conclusion we can made from  this comparison is the important role played by the boundary conditions on rigid boundaries. Almost all experiments show that the model is the most flexible with respect to control of coefficients $\alpha$, this control allows us to bring the model's solution  closer to the solution of the high-resolution model or to the observed data.   In the experiment with the fourth order model in the square box, it is the control of the numerical scheme near the boundary that  represents the major part of the total flexibility of the model (see \rfg{evol-xi.sq} on the right). 

Optimal $\alpha$ found in the assimilation window remain optimal long time after the end of assimilation improving the forecasting ability of the model. We could see that the fourth order model in the square box allows us to divide by two  the forecast error of the  20 days forecast. Optimal $\alpha$ obtained in one month assimilation remains optimal even for a one year run of the Black sea model. 

  Finally, the  long time sensitivity of the model's solution to $\alpha$ exceeds the sensitivity to almost all other parameters including the sensitivity to initial conditions. A perturbation of $\alpha$ of a given small norm results in a bigger perturbation of the model's solution than a perturbation of some other parameter of an equal norm. 
 
   However, we could see that the influence of boundary conditions  is only important on long time scales, i.e. time scales that exceeds the characteristic time of the domain. In both experiments presented above the characteristic  time was approximately equal to 5 days (as it was mentioned above, we use $T_{char}=L/\sqrt{gH_0}$), and in both experiments the sensitivity to $\alpha$ becomes important on scale longer than 5 days. On the other hand, on short scales, ($T<0.1 T_{char}$)  it is the bottom topography that influences the most the model's solution. Both in the Black sea and in the square box the sensitivity to topography is approximately 40 times more important than the sensitivity to $\alpha$. 
   
In addition to that, we should note that usually prescribed boundary conditions (impermeability and no-slip conditions have been used here as the initial guess for the cost function  the minimization) seem not to be optimal for    the model. As we can see in \rfg{evol-xi.sq} and in \rfg{blk}, modifying $\alpha$ we can bring the model much closer to the high resolution model or to the observational data. Optimal numerical scheme allows to divide  the cost function's value by 4 in the experiment with the Black sea model and by 3.5 in the experiment with the fourth order model in a square. But, the numerical scheme is modified in the assimilation process.    As it has been shown  in \cite{sw-nl}, optimal numerical scheme near the boundary may violate even impermeability condition indicating the necessity to change the domain's geometry.

Taking into account an important influence of the numerical scheme that introduces boundary conditions into the model,  it is reasonable to think about identification of the  optimal scheme by data assimilation process instead of prescribing classical boundary conditions.

{\bf Acknowledgments. } The author thanks  Gennady Korotaev from   Marine Hydrophysical Institute, National Academy of Sciences of
Ukraine for providing the model parameters and data for the upper layer model of the Black Sea. 
 

\mkpicstoend 

\end{document}